\documentclass[twocolumn,prl,10pt,superscriptaddress]{revtex4}
\pdfoutput=1
\usepackage{graphicx}
\usepackage{graphics}
\usepackage{floatflt}
\graphicspath{{figures/}{figures2/}}
\DeclareGraphicsExtensions{.eps,.png}

 \newcommand{\ket}[1]{$| #1 \rangle$}
 \newcommand{\bra}[1]{$\langle #1 |$}
 \newcommand{\extcjj}{$\Phi_x^{cjj}$}
 
 \newcommand{\altwoothree}{Al$_2$O$_3$}
 \newcommand{\micron}{$\mu$m }
 \newcommand{\siotwo}{SiO$_2$}
\newcommand{\length}{$L$}
\newcommand{\width}{$w$}

\begin{document}
\title{Geometrical dependence of low frequency noise in superconducting flux qubits}
\author{T. Lanting}
\email{tlanting@dwavesys.com}
\author{A.J. Berkley}
\affiliation{D-Wave Systems Inc., 100-4401 Still Creek Drive, Burnaby, B.C., V5C 6G9, Canada}
\author{B. Bumble}
\affiliation{Jet Propulsion Laboratory, California Institute of Technology, Pasadena CA, USA}
\author{P. Bunyk}
\affiliation{D-Wave Systems Inc., 100-4401 Still Creek Drive, Burnaby, B.C., V5C 6G9, Canada}
\author{A. Fung}
\affiliation{Jet Propulsion Laboratory, California Institute of Technology, Pasadena CA, USA}
\author{J. Johansson}
\affiliation{D-Wave Systems Inc., 100-4401 Still Creek Drive, Burnaby, B.C., V5C 6G9, Canada}
\author{A. Kaul}
\author{A. Kleinsasser}
\affiliation{Jet Propulsion Laboratory, California Institute of Technology, Pasadena CA, USA}
\author{E. Ladizinsky}
\author{F. Maibaum}
\author{R. Harris}
\author{M.W. Johnson}
\author{E. Tolkacheva}
\author{M. H. S. Amin}
\affiliation{D-Wave Systems Inc., 100-4401 Still Creek Drive, Burnaby, B.C., V5C 6G9, Canada}

\date{\today}

\begin{abstract}
A general method for directly measuring the low-frequency flux noise (below 10 Hz) in compound Josephson junction superconducting flux qubits has been used to study a series of 85 devices of varying design.  The variation in flux noise across sets of qubits with identical designs was observed to be small. However, the levels of flux noise systematically varied between qubit designs with strong dependence upon qubit wiring length and wiring width. Furthermore, qubits fabricated above a superconducting ground plane yielded lower noise than qubits without such a layer. These results support the hypothesis that localized magnetic impurities in the vicinity of the qubit wiring are a key source of low frequency flux noise in superconducting devices.
\end{abstract}

\maketitle
    %

    %

    %




Qubits implemented in superconducting integrated circuits show considerable promise as building blocks of scalable quantum processors. However, low frequency noise in superconducting devices places fundamental limitations on their use in quantum information processing~\cite{koch1983, yoshihara2006,wellstood1987:lowfrequency,wellstood1987:excess}. Recent theoretical work has highlighted several potential sources for low frequency noise. These include ensembles of two level systems (TLS) that could be associated with dielectric defects~\cite{simmonds2004,martinis2005,koch2007}, magnetic impurities in surface oxides on superconducting wiring~\cite{ioffe2008} and flux noise induced by spin flips at dielectric interfaces~\cite{desousa2007}. Characterizing low frequency noise is an essential step in understanding its mechanism and in developing fabrication strategies to minimize its amplitude. Several techniques have been exploited to indirectly measure low frequency noise in superconducting qubits~\cite{harris2007,bialczak2007}. This article describes a technique for directly measuring low frequency noise in RF-SQUID flux qubits. We present measurements performed on a series of qubits of varying wiring lengths and widths and qubits with and without superconducting shielding layers.

\begin{figure}
  \includegraphics[width=3.5in]{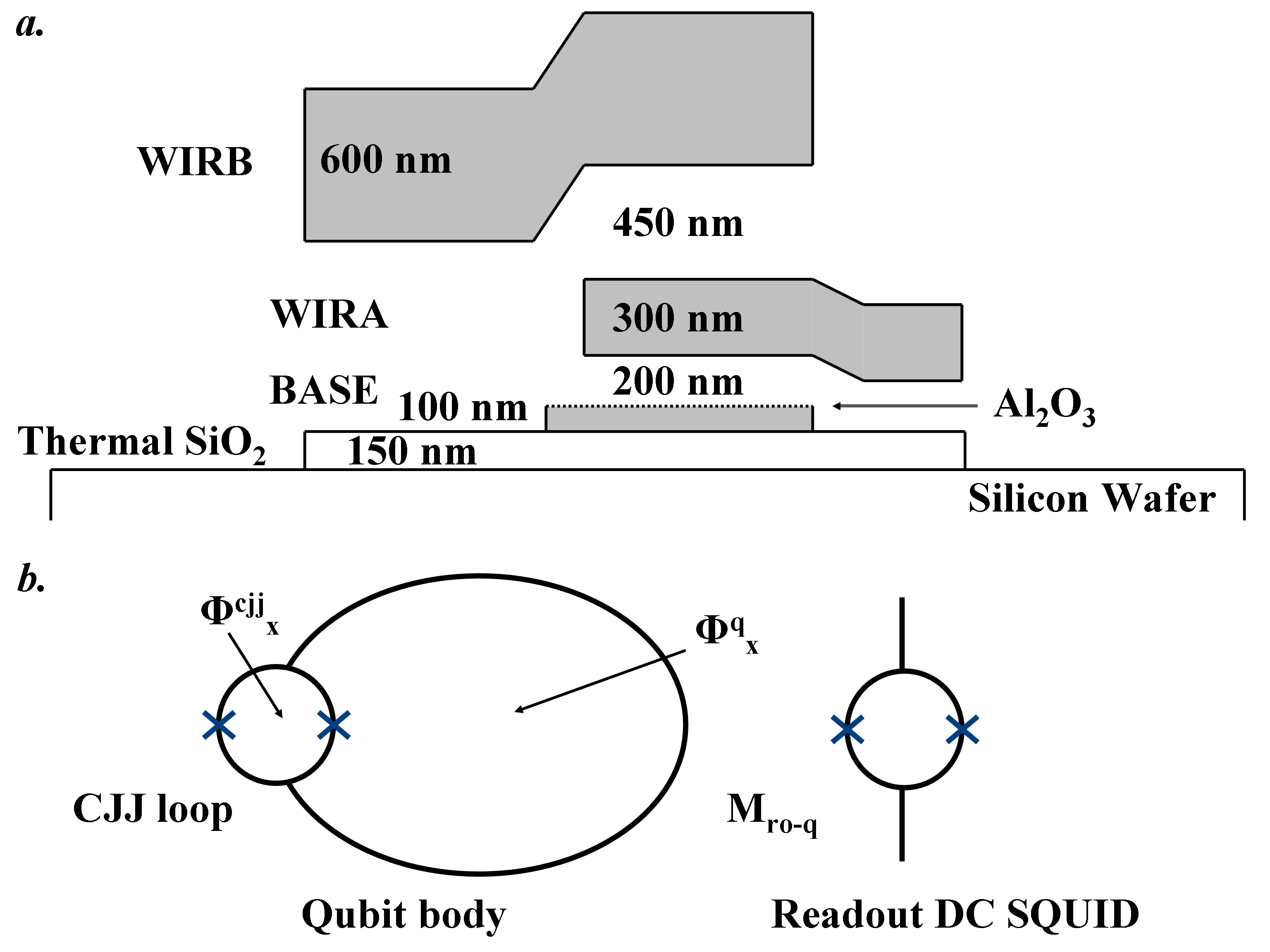}
  \caption{{\bf{a.}} Cross-section of the fabrication stack. There are three Nb metal layers, BASE, WIRA, and WIRB. The trilayer is grown on the top of the BASE layer. Wiring layers are insulated from each other with layers of sputtered \siotwo. {\bf{b.}} CJJ RF-SQUID qubit schematic.
  \label{FIG:cjj}}
\end{figure}

The devices described in this paper were fabricated on an oxidized Si wafer with a Nb/Al/\altwoothree/Nb trilayer process. There were two additional wiring layers, WIRA, and WIRB, above the trilayer (see Fig \ref{FIG:cjj}a). All wiring layers were insulated from each other with layers of sputtered \siotwo~. Eighty-five qubits with a range of geometries (wiring length, wiring width, and the presence or absence of shielding planes) were tested. Qubit wiring lengths ranged from 350 \micron to 2.1 mm and wiring widths ranged from 1.4 \micron to 3.5 $\mu$m. Moreover, the qubits were drawn from several wafers to control for variability in fabrication process conditions.

The compound Josephson junction (CJJ) RF-SQUID is shown schematically in Fig. 1b and consists of a small CJJ loop (inductance $\sim 10$ pH) and the main qubit loop (inductances vary from $\sim 80$ to $\sim 800$ pH). The loops are externally flux biased with $\Phi_x^{cjj}$ and $\Phi_x^q$, respectively. The CJJ loop contains two Josephson junctions with critical current $I^q_c$ when connected in parallel. For $L_{cjj} \ll L_q$ the Hamiltonian for an isolated device can be approximately expressed as ~\cite{harris2008}:

\begin{equation}
H_{\text{rf}}(\Phi^q,Q) = \frac{1}{2C^q}Q^2 + U(\Phi^q),
\label{EQN:fullHamiltonian}
\end{equation}

\begin{equation}
U(\Phi^q) = \frac{(\Phi^q - \Phi^q_x)^2}{2L^q} - E_J \cos\left[\frac{\pi \Phi_x^{cjj}}{\Phi_0}\right]
\cos\left[\frac{2\pi\Phi^q}{\Phi_0}\right],
\end{equation}
where $\Phi^q$ is the total flux, $Q$ is the charge stored in the net capacitance $C^q$ across the junctions, $E_J = \Phi_0 I^q_c/2\pi$ and $\Phi_0 = h/2e$. The potential energy $U(\Phi_q)$ is monostable when $\beta = 2\pi L_q I_c \cos( \pi \Phi_x^{cjj}/\Phi_0)/\Phi_0 < 1$ and classically bistable, with two counter-circulating persistent current states (denoted as \ket{0} and \ket{1}) possessing persistent current of magnitude $| I_p | = |$\bra{n} $\Phi^q/L^q$ \ket{n}$|$ for $\beta > 1$. Ignoring all but the two lowest energy levels (two level approximation), one can map an isolated CJJ RF-SQUID onto a qubit Hamiltonian:

\begin{equation}
H_{\text{q}} = -\frac{1}{2} [\epsilon \sigma_z + \Delta \sigma_x],
\end{equation}
where $\epsilon =  2| I_p |\Phi_x^q$ is the energy bias, $\Delta$ is the tunneling energy between \ket{0} and \ket{1}, and $\sigma_z, \sigma_x$ are Pauli matrices. For an RF-SQUID qubit both $|I_p|$ and $\Delta$ are functions of \extcjj. A DC SQUID, inductively coupled to the qubit with mutual inductance $M_{\text{ro-q}}$, distinguishes between the qubit states by measuring the flux generated by the persistent current for \extcjj $= n \Phi_0$ where $\Delta \approx 0$.

Our qubit magnetometry technique consists of initializing the qubit at its degeneracy point (\extcjj  $ = \Phi_0/2$, $\Phi^q = 0$), raising the tunnel barrier by applying a 30 $\mu$s linear ramp from \extcjj $= \Phi_0/2$ to \extcjj $= \Phi_0$ to localize the qubit in \ket{0} or \ket{1}, and then measuring its state with the DC-SQUID. We repeat this bias and measure cycle $n$ times, assigning each \ket{0} measurement a value 0 and each \ket{1} measurement a value 1. We then average these $n$ measurements to produce a population measurement $P \in [0,1]$ which has an uncertainty $1/\sqrt{4n}$. In the absence of flux noise one would expect $P = 0.5$. However, in the presence of low frequency flux noise, the population measurements will fluctuate about $P = 0.5$ with an amplitude that depends on the noise scale. We can thus use $P$ to directly probe the flux environment experienced by the qubit as a function of time.

\begin{figure}
  \includegraphics[width=3.5in]{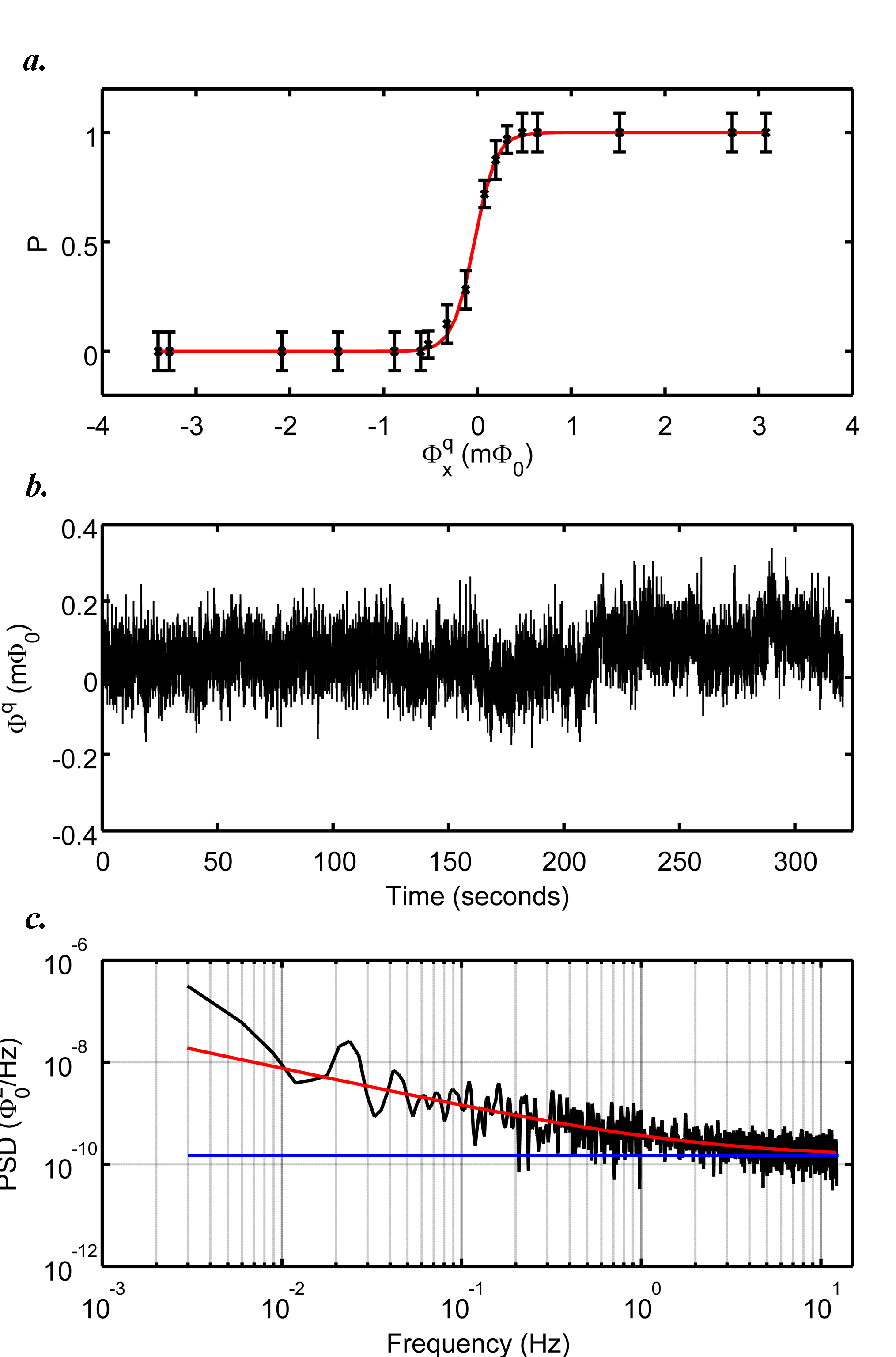}
  \caption{Flux noise measurements for a typical qubit. {\bf{a.}} Static qubit population $P$ vs. external flux bias. The fit is shown in red. {\bf{b.}} Qubit flux versus time. {\bf{c.}} Power spectral density of flux signal. The fit to Eq. (\ref{eqn:fitfunction}) is shown in red and the white noise level expected from a binomial distribution is shown in blue.}
  \label{FIG:oberon-single-qubit}
\end{figure}

To calibrate the mapping from $P$ to flux noise amplitude we first measured the flux periodicity of the RF-SQUID. We then assume a phenomenological form for population versus external flux bias:
\begin{equation}
P(\Phi^q_x,t) =  \frac{1}{2} \left[1 + \tanh\left(\frac{\Phi^q_x + \Phi_n(t) - \Phi^q_0}{2\delta}\right) \right]
\label{eqn:pop-tanh}
\end{equation}
where $\Phi_n(t)$ is a time dependent flux noise, $\Phi^q_0$ is the external flux needed to balance $P=0.5$ ($\epsilon = 0$), and $\delta$ captures the breadth of the transition. With sufficient averaging, we measure a static population distribution $\overline{P(\Phi^q_x)} = \frac{1}{2}\left[1+\tanh(\frac{\Phi^q_x - \Phi^q_0}{2\delta})\right]$ which allows us to calibrate $\Phi^q_0$ and $\delta$ (see Fig. \ref{FIG:oberon-single-qubit}a). Inverting Eq. \ref{eqn:pop-tanh} then allows one to convert measurements of $P$ to $\Phi_n(t)$.

Figure \ref{FIG:oberon-single-qubit} shows example qubit magnetometry measurements. Each flux measurement is derived from 128 population measurements. Time traces were collected for 325 seconds. The white noise level of the resulting PSD is consistent with the error expected from a binomial distribution of population measurements. At a fixed temperature this noise level can be reduced by sampling at a higher frequency. In practice, we are limited to a sample frequency of 5 kHz by the bandwidth of the control lines connecting the qubit to the room temperature electronics.

\paragraph{}
In addition to the aforementioned white noise, a typical PSD also shows a $1/f$ frequency dependence for $f < 10$ Hz. To separate the contribution to the PSD from white noise and $1/f$ noise, we define the $1/f$ contribution as
\begin{equation}
S_{\Phi}(f) = A \left(\frac{1\ \rm{Hz}}{f}\right)^{-\alpha}
\end{equation}
and fit the measured PSD from every qubit tested with the function:
\begin{equation}
\label{eqn:fitfunction}
PSD(f) = S_{\Phi}(f) + w_n,
\end{equation}
where $A$ represents the magnitude of the $1/f$ noise at 1 Hz ($\Phi_0^2$/Hz), $\alpha$ represents the power of the frequency dependence, and $w_n$ represents the white noise level. We interpret the $1/f$ power extracted from the PSD fit as a low frequency flux noise signal $S_{\Phi}$ that biases the qubit~\footnote{Another potential source of noise in our devices are fluctuations in the junctions themselves. We have independently estimated the level of low frequency critical current noise in the qubit junctions. We can place an upper limit of $\sqrt{S_{I_c}(\text{1 Hz})} = 3 \times 10^{-6}$ at 1 Hz on relative $I_c$ fluctuations in our junctions. The flux noise we are measuring is too large to be explained by $I_c$ fluctuations.}. Measurements of the current noise of the room temperature current sources that provide $\Phi^q_x$ reveal that their low frequency noise is over a factor of 50 below the smallest measured $S_\Phi$. The fit values of $\alpha$ were clustered around $\alpha = 1.00 \pm 0.15$.

\begin{figure}
  \includegraphics[width=3.5in]{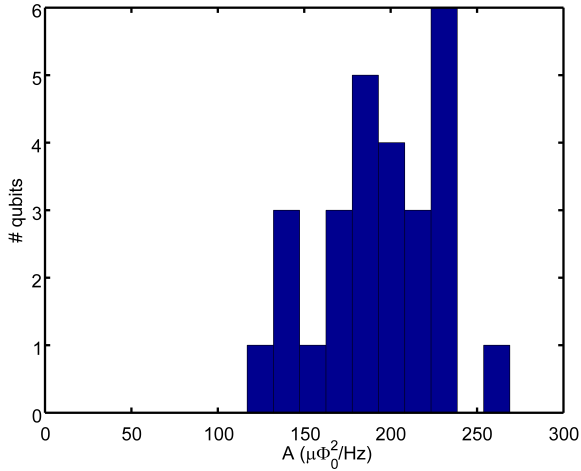}
  \caption{Variation of qubit flux noise across 27 identical qubits.}
  \label{FIG:leda}
\end{figure}

\begin{figure}
  \includegraphics[width=3.5in]{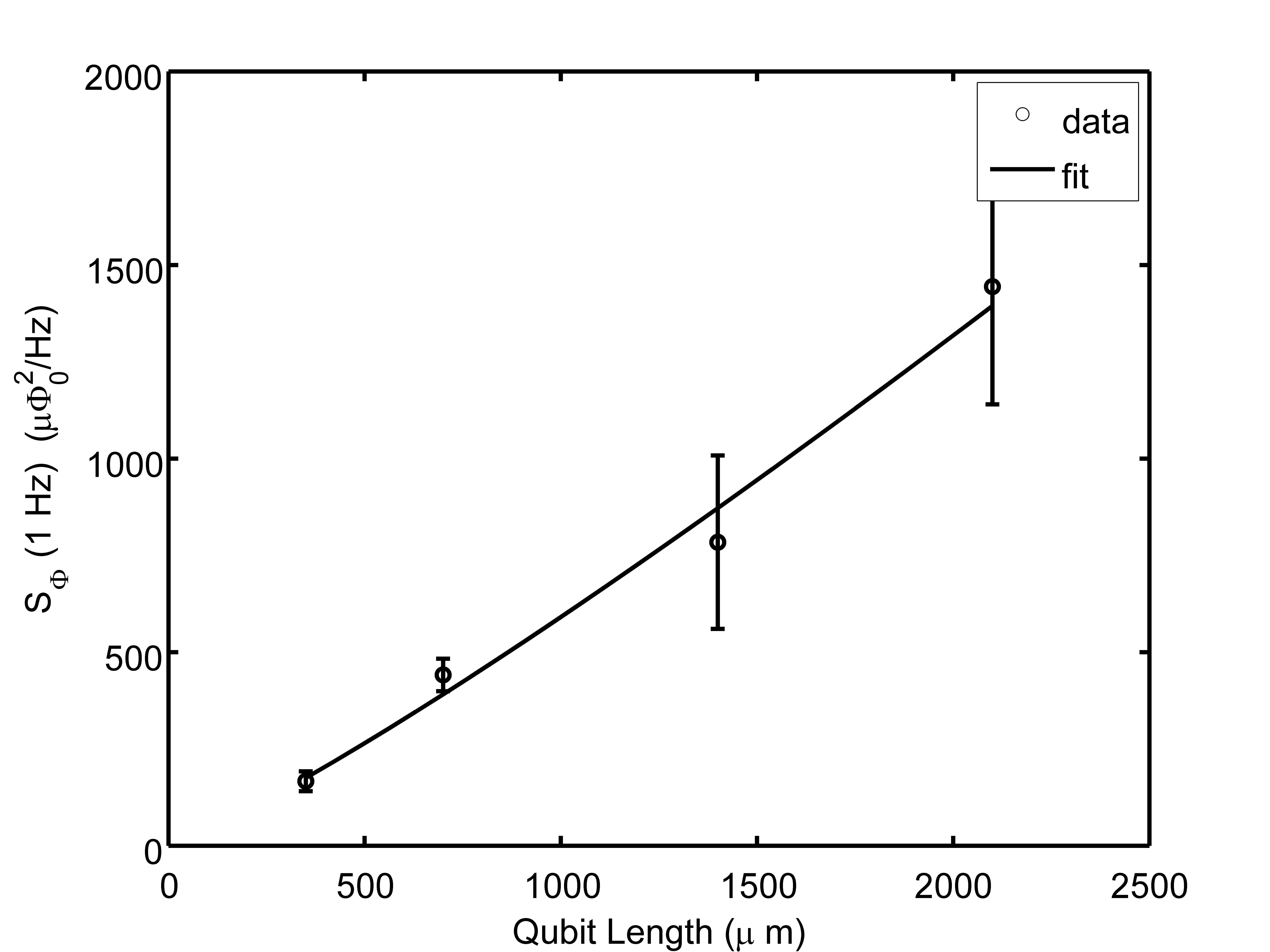}
  \caption{Qubit noise for different wiring lengths and a shielding plane under the qubit.}
  \label{FIG:length-dep}
\end{figure}

\begin{figure}
  \includegraphics[width=3.5in]{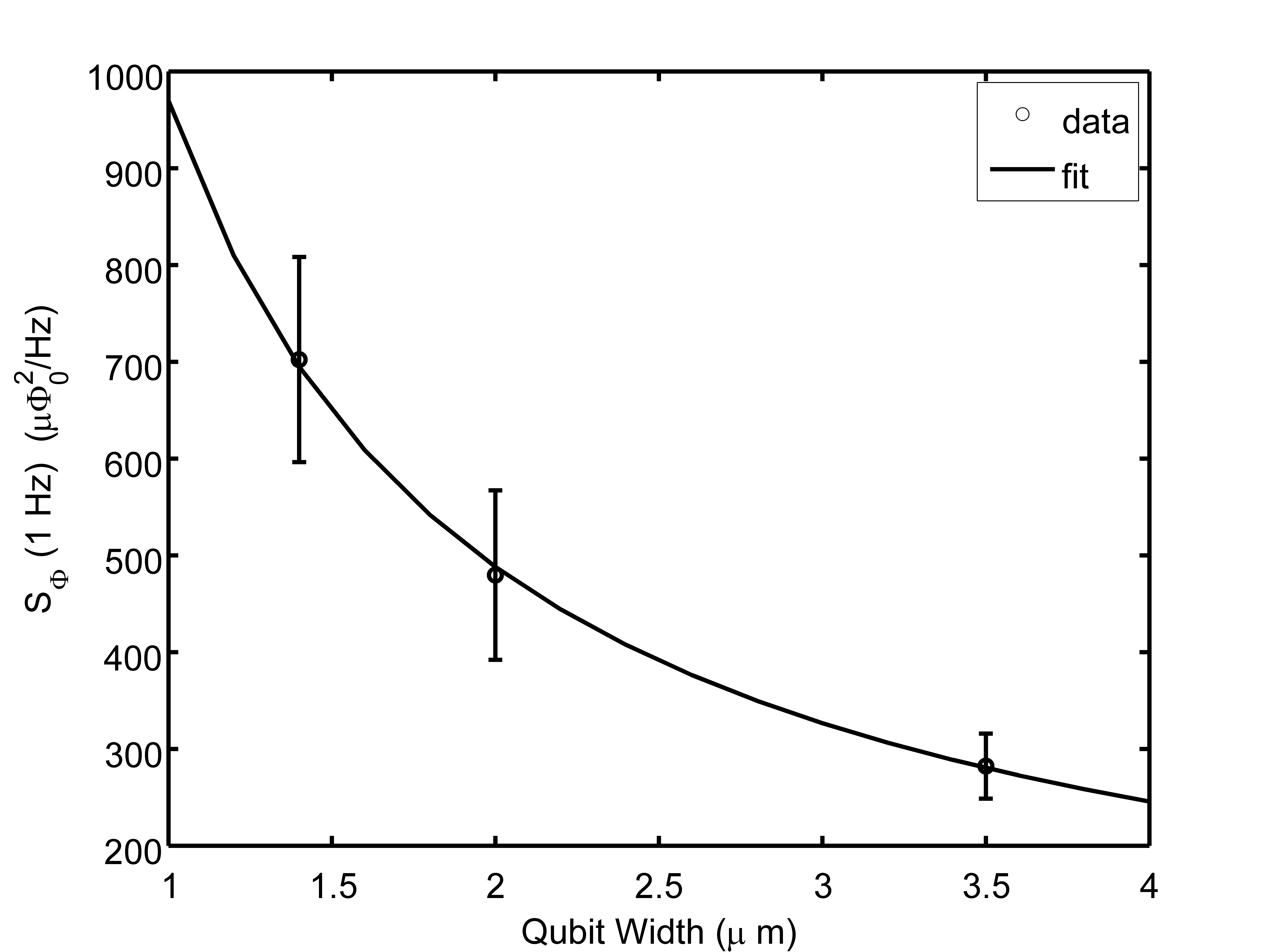}
  \caption{Qubit noise for different wiring widths, no ground plane present.}
  \label{FIG:width-dep}
\end{figure}

\paragraph{}
To investigate the variation in low frequency noise between qubits of identical design we measured a set of 27 identical qubits (700$\mu$m long, 1.4$\mu$m wide). Figure \ref{FIG:leda} shows a histogram of the low frequency flux noise $S_\Phi (\text{1 Hz})$ for these qubits. We measure a median $A = 1.9\pm 0.3 \times 10^{-10} \Phi_0^2/\text{Hz}$.

\paragraph{}
In figure \ref{FIG:length-dep} we compare the value of $A$ as a function of wiring length for qubits that are $1.4\ \mu$m wide and have a shielding plane under the qubit wiring. A fit of the form $S_{\Phi} ( \text{1 Hz}) = A_0 L^\beta$, yielded $A_0 = 1.7\pm 0.3 \times 10^{-10}\Phi_0^2/$Hz and $\beta = 1.14 \pm 0.15$. $A_0$ measures the 1/f power scaled to a length of $350\ \mu$m and $\beta$ measures the scaling with length. The $1/f$ power scales approximately linearly with the wiring length of the qubit. Performing the same comparison for qubits that are 3.5 \micron wide and have no shielding plane yields  $A_0 = 9.5 \pm 0.3 \times 10^{-11}\Phi_0^2/$Hz and $\beta = 1.1\pm 0.25$.

\paragraph{}
In figure \ref{FIG:width-dep} we compare the $1/f$ noise as a function of wiring width in qubits that are 350 \micron long and have no shielding plane. We see a dependence on the $1/f$ amplitude with qubit width. A fit of the results to the form $S_{\Phi} (\text{1 Hz}) = B_0 w^{\gamma}$,
where $B_0$ measures the $1/f$ power scaled to a width of 1\micron and $\gamma$ measures the scaling with width. The fit yielded $B_0 = 9.6 \pm 0.5 \times 10^{-10} \Phi_0^2$/Hz and $\gamma = -0.98\pm 0.10$. The $1/f$ power scales approximately as the inverse of the wiring width of the qubit.

\paragraph{}
There are clear differences in qubits that are geometrically identical except for the presence or absence of a ground plane underneath the device. Table \ref{tbl:shieldnoshield} shows a comparison of the measured noise between such pairs of qubits. In all cases, qubits with a ground plane underneath the device were quieter by 30 to 50$\%$.

\begin{table}
\caption{\label{tbl:shieldnoshield} Comparison of qubits with and without BASE shielding}
\begin{tabular}{|c|c|c|c|}
\hline
 Description &  Shield & $\sqrt{S_{\Phi} (\text{1 Hz})}\ (\mu\Phi_0/\sqrt{\rm Hz})$ \\
\hline
WIRA, \length=2.1 mm, \width=3.5 $\mu$m & Yes & $17\pm 1$ \\
WIRA, \length=2.1 mm, \width=3.5 $\mu$m & No & $25\pm 4$ \\
\hline
WIRB, \length=2.1 mm, \width=1.4 $\mu$m & Yes & $29\pm 2$\\
WIRB, \length=2.1 mm, \width=1.4 $\mu$m & No & $34\pm 4$\\
\hline
WIRA, \length=1.4 mm, \width=3.5 $\mu$m & Yes & $15\pm 2 $\\
WIRA, \length=1.4 mm, \width=3.5 $\mu$m & No & $ 21 \pm 2 $\\
\hline
WIRA, \length=1.4 mm, \width=1.4 $\mu$m & Yes & $21\pm 2$ \\
WIRA, \length=1.4 mm, \width=1.4 $\mu$m & No & $27\pm 2$ \\
\hline
WIRB, \length=1.4 mm, \width=1.5 $\mu$m & Yes & $21\pm 2$ \\
WIRB, \length=1.4 mm, \width=1.5 $\mu$m & No & $27 \pm 3$ \\
\hline
\end{tabular}
\end{table}

\paragraph{}
Qubit magnetometry on this series of devices has revealed a clear dependence of the low frequency noise on geometry. The scaling with wiring length and width suggests a noise source local to the qubit wiring. For example, the microscopic model of low frequency noise described in ~\cite{ioffe2008,sendelbach2008:magnetism} suggests low frequency noise due to localized but interacting magnetic moments (the most likely candidate being defects in Nb$_2$O$_5$ layers on the qubit wiring) where the power should scale as the ratio of qubit length to qubit width. For an areal defect density of $\sigma_s$, the magnitude of the noise in this model is estimated to be
\begin{equation}
S_{\Phi} \sim \frac{4}{\pi}(\mu_0 \mu_B)^2 \sigma_s \frac{L}{w} \frac{1}{f}
\label{EQN:ioffe}
\end{equation}
for a qubit of length $L$ and width $w$. For a 350 \micron long, 1.4 \micron wide qubit and $\sigma_s = 10^{16}/\rm{m}^2$~\cite{ioffe2008}, Eq (\ref{EQN:ioffe}) predicts $S_{\Phi} (\text{1 Hz}) \sim 10^{-10} \Phi_0^2/$Hz which is in rough agreement with the results reported herein. Moreover, the measured geometrical dependence corroborates the predicted functional form of this model.
However, it should be noted that this model is only appropriate for a qubit in free space. The current density in a superconducting wire in free space is peaked at the edges of the wire~\cite{vanduzerbook}. The presence of a shielding plane close to the qubit loop makes the current distribution more uniform and should thus decrease the effect of local magnetic impurities~\cite{ioffe2008}. This hypothesis is supported by the clear reduction in low frequency noise in the presence of a ground plane under the qubit wiring.

\paragraph{}
A similar reduction in the low frequency noise should be observed when the qubits are shielded by a plane either above or below the qubit wiring. However, directly comparing qubits with shielding under and over their wiring reveals a clear difference. The data in Table \ref{tbl:basevssky} show that qubits with shielding under their wiring are systematically quieter than qubits with shielding over their wiring.  This could be due to the increased separation between qubit wiring and the shielding layer above the qubit as compared to the separation between the qubit and the shielding layer below it (see figure \ref{FIG:cjj}). In addition, the current distribution in qubits with shielding layers is predominantly along the wiring surface closest to the shielding layer.  For qubits with underlying shielding layers, current flow is distributed along the bottom surface of the qubit wiring. This surface is naturally protected from subsequent fabrication and ambient conditions and should exhibit fewer impurities. Note also that shielding under the qubit wiring isolates the qubit from the Si/\siotwo$\ $interface on the substrate. Impurities at this interface could be responsible for coupling flux into the qubit ~\cite{desousa2007}.

\begin{table}
\caption{\label{tbl:basevssky} Comparison of qubits with shielding above and below the qubit wiring}
\begin{tabular}{|c|c|c|c|}
\hline
 Description &  Shield & $\sqrt{S_{\Phi} (\text{1 Hz})}\ (\mu\Phi_0/\sqrt{\rm Hz})$ \\
\hline
 WIRA, \length=0.7 mm, \width=1.4 \micron & under & $18 \pm 1$ \\
 WIRA, \length=0.7 mm, \width=1.4 \micron & over & $21 \pm 1$ \\
\hline
 WIRA, \length=2.1 mm, \width=1.4 \micron & under & $26 \pm 2$ \\
 WIRA, \length=2.1 mm, \width=1.4 \micron & over & $38 \pm 2$ \\
\hline
 WIRA, \length=1.4 mm, \width=1.4 \micron & under & $21\pm 2$ \\
 WIRA, \length=1.4 mm, \width=1.4 \micron & over & $28 \pm 2$ \\
\hline
\end{tabular}
\end{table}

\paragraph{}
The magnetometry technique described in this paper is an effective way of probing low frequency flux noise in superconducting flux qubits. This technique revealed that qubits having varying lengths, widths, and shielding are subject to systematically different levels of noise. The behavior of the measured flux noise is in agreement with theoretical microscopic models that postulate that the source of flux noise is magnetic impurities proximal to the qubit wiring. However, the measurements do not rule out local impurities in the dielectric insulating layers. These measurements suggest that to reduce low frequency noise in superconducting qubits, the ratio of qubit length to width should be reduced as much as possible. A shielding layer close to the qubits and preferably between the qubit wiring and the Si/\siotwo$\ $ interface further reduces flux noise.

We thank J. Hilton, G. Rose, G. Dantsker, C. Rich, E. Chapple, P. Spear, F. Cioata, B. Wilson and F. Brito for useful discussions. Samples were fabricated by the Microelectronics Laboratory of the Jet Propulsion Laboratory, operated by the California Institute of Technology under a contract with NASA.

\bibliography{master_references}

\end{document}